\documentclass[twocolumn,showpacs,superscriptaddress,amsmath,amssymb,aps,pre]{revtex4}
\usepackage{epsfig}
\usepackage{yfonts}
\usepackage{bm}% bold math
\usepackage{amssymb}
\usepackage{graphicx}
\usepackage{color}
\usepackage{wrapfig,subfigure}
\usepackage{pst-grad} % For gradients
\usepackage{pst-plot}
\swabfamily

\newcommand{\ep}{\varepsilon}

\def\p12{p_{12}({\bf q},t)}

\begin{document}

\title{An Ideal Mean-Field Transition in a Modulated Cold Atom System}

\author{Myoung-Sun Heo}
\affiliation{%
Department of Physics and Astronomy, Seoul National University, Seoul 151-747, Korea
}%
\author{Yonghee Kim}
\affiliation{%
Department of Physics and Astronomy, Seoul National University, Seoul 151-747, Korea
}%
\author{Kihwan Kim}
\affiliation{%
Department of Physics and Astronomy, Seoul National University, Seoul 151-747, Korea
}%
\author{Geol Moon}
\affiliation{%
Department of Physics and Astronomy, Seoul National University, Seoul 151-747, Korea
}%
\author{Junhyun Lee}
\affiliation{%
Department of Physics and Astronomy, Seoul National University, Seoul 151-747, Korea
}%
\author{Heung-Ryoul Noh}
\affiliation{%
Department of Physics, Chonnam National University, Gwangju 500-757, Korea
}%
\author{M. I. Dykman}
\email{dykman@pa.msu.edu}
\affiliation{%
Department of Physics and Astronomy, Michigan State University, East Lansing, Michigan 48824, USA
}%
\author{Wonho Jhe}
\email{whjhe@snu.ac.kr}
\affiliation{%
Department of Physics and Astronomy, Seoul National University, Seoul 151-747, Korea
}%

\date{\today}
\begin{abstract}
We show that an atomic system in a periodically modulated optical trap displays an ideal mean-field symmetry-breaking transition.  The symmetry is broken with respect to time translation by the modulation period. The transition is due to the interplay of the long-range interatomic interaction and nonequilibrium fluctuations. The observed critical dynamics, including anomalous fluctuations in the symmetry broken phase, are fully described by the proposed microscopic theory.
\end{abstract}

\maketitle

The mean-field approach has been instrumental for developing an insight into symmetry breaking transitions in thermal equilibrium systems \cite{Kadanoff2009}. It has been broadly applied also to nonequilibrium systems, the studies of pattern formation being an example \cite{Cross2009}. However, close to the phase transition point the mean field approximation usually breaks down. This happens even in finite-size systems provided they are sufficiently large.

In this paper we study a nonequilibrium system of $\sim 10^7$ particles which, as we show, displays an ideal mean-field symmetry breaking transition. It is accompanied by anomalous fluctuations, which are also described by a mean-field theory. The system is formed by moderately cold atoms in a magneto-optical trap (MOT) \cite{Walker1990,Sesko1991}. The atoms are periodically modulated in time \cite{Kim2003}. Periodically modulated systems form one of the most important classes of nonequilibrium systems, both conceptually and in terms of applications. They have discrete time-translation symmetry: they are invariant with respect to time translation by modulation period $\tau_F$. Nevertheless, they may have stable vibrational states with periods that are multiples of $\tau_F$, in particular $2\tau_F$. Period doubling is well-known from parametric resonance, where a system has two identical vibrational states shifted in phase by $\pi$. It is broadly used in classical and quantum optics and has attracted significant attention recently in the context of nano- and micro-mechanical resonators \cite{Turner1998,Buks02,Chan2007,Lifshitz2003}.

In a many-body system, dynamical period doubling in itself does not break the time translation symmetry. This is a consequence of fluctuations. Even though each vibrational state has a lower symmetry, fluctuations make the states equally populated and the system as a whole remains symmetric. However, if as a result of the interaction the state populations become different, the symmetry is broken. It is reminiscent of the Ising transition where the interaction leads to preferred occupation of one of the two equivalent spin orientations, except that the symmetry is broken in time. For atoms in a parametrically modulated MOT spontaneous breaking of the discrete time-translation symmetry was observed experimentally by Kim {\it et al.} \cite{Kim2006}.

Here we show that the time-translation symmetry breaking in a modulated MOT results from the cooperation and competition between the inter-particle interaction and the fluctuations that lead to atom switching between the vibrational states. The interaction is weak. On its own, it cannot change the state populations. However, it may become strong enough to change the rates of fluctuation induced switching, which in turn will cause the population change. We provide a quantitative theory of the phenomenon. We measure the critical exponents and the frequency dispersion of the susceptibility. The observations are consistent with the mean-field behavior and are in good agreement with the theory.

In the experiment, $^{85}$Rb atoms in a MOT were cooled down to $\approx$0.4 mK. Full three-dimensional confinement was achieved with three pairs of counter-propagating laser beams. The intensity of the beams along the MOT axis $z$ was 0.19~mW/cm$^2$, the  transverse beam intensities were 5 to 10 times larger. The magnetic field gradient at the trap center was 10~G/cm. The transverse beams were detuned from the atomic transition by $\delta \approx -2.3\Gamma_p$, the longitudinal beams were further detuned by 5~MHz (the atomic decay rate is $\Gamma_p/2\pi\approx 5.9$~MHz). The atomic cloud motion was essentially one-dimensional, along $z$-axis, with frequency $\omega_0/2\pi \approx 45$~Hz and the damping rate $\Gamma \approx 36$ s$^{-1}$. The total number of trapped atoms $N_{\rm tot}$ was varied by changing the hyperfine-repumping laser intensity at a decrease rate of 0.5\% per second. For the order parameter and the variance measurements very close to criticality, the decrease rate was reduced to 0.03\%.

\begin{figure}[h]
  %\centering
  \includegraphics[scale=0.7]{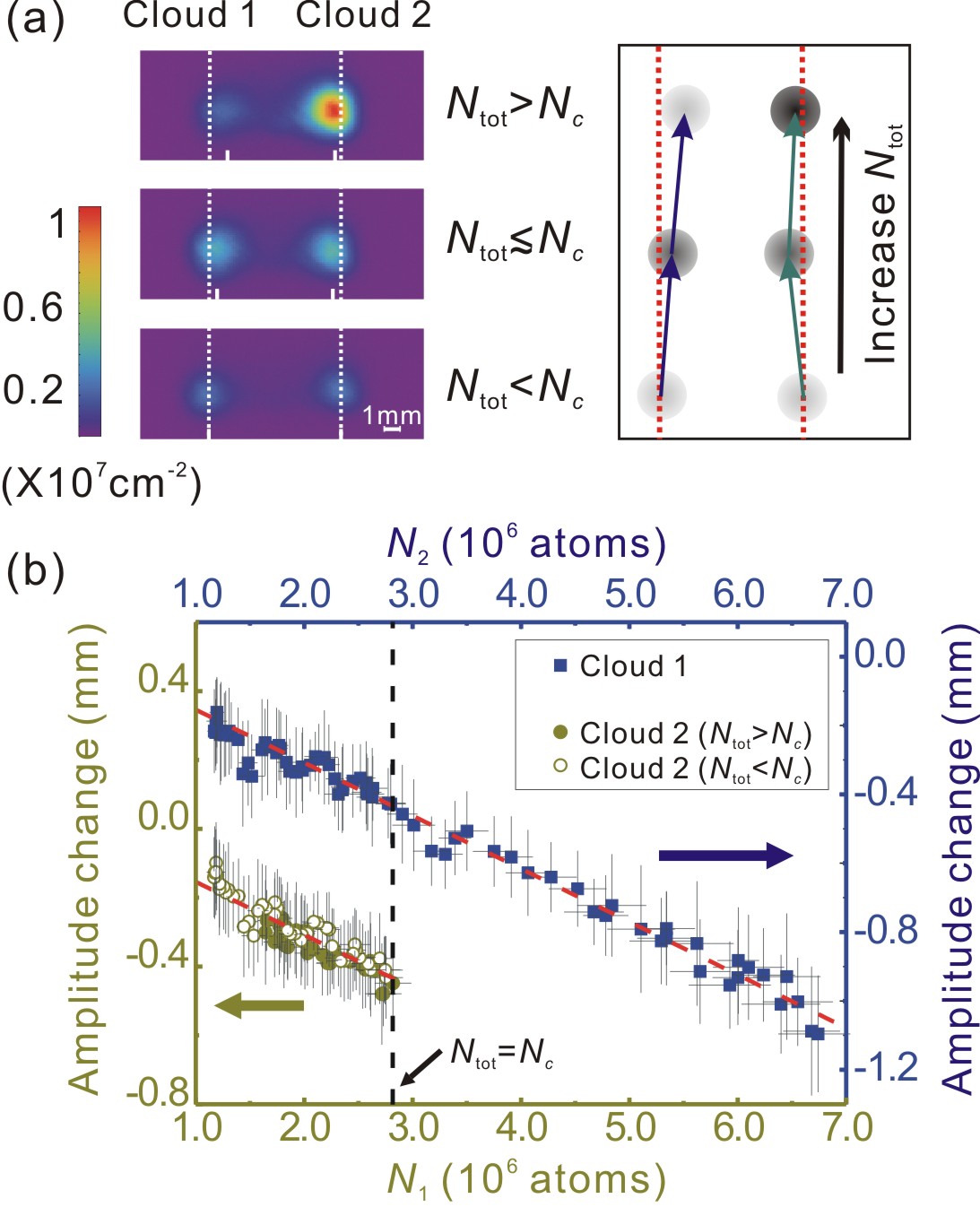}
  \caption{(Color) (a) The two-dimensional density profiles of the vibrating clouds for the maximal displacement along the MOT axis ($\approx$ 4.5 mm). The lower, middle, and upper panels refer to the symmetric, close to critical, and broken-symmetry states, with $N_{\rm tot}\approx1.5\times10^6$, $5.2\times10^6$, and $6.7\times10^6$ , respectively; $N_c \approx 5.6\times 10^6$. The cloud centers are indicated by the ticks at the bottom of the panels and are sketched in the right panel. (b) The amplitude changes of the cloud vibrations as functions of the population of the other cloud. The red dashed lines show the expected linear dependence. The black dashed line shows $N_1=N_2=N_c/2$. The amplitude of the cloud with smaller population beyond the transition, cloud 1, monotonically increases with $N_{\rm tot}$, whereas that of the larger cloud, cloud 2, is maximal at criticality. The error bars show the standard deviation for 20 measurements.}\label{fig1}
\end{figure}

The beam intensities were modulated at frequency $\omega_F=2\omega_0$ by acousto-optic modulators. When the modulation amplitude exceeded a threshold value, after a transient, the atomic cloud split into two clouds \cite{Kim2003}, which were vibrating in counter-phase with frequency $\omega_F/2$, see Fig.~\ref{fig1}a. We took snapshots at the maximum separation of the clouds at the frame rate of 1-50 Hz to obtain the population of each cloud. For a small total number of trapped atoms $N_{\rm tot}$ the populations were equal and the cloud vibrations had equal amplitudes.

Once $N_{\rm tot}$ exceeded a critical value $N_c$ ($N_c\sim 10^7$ in our experiment), the average populations of the clouds became different, as observed earlier \cite{Kim2006}. The vibration amplitudes became different, too, see Fig.~\ref{fig1}b. This is spontaneous breaking of the discrete time-translation symmetry, the system becomes invariant with respect to time translation by $2\tau_F$, not the modulation period $\tau_F$.

As phase transitions in equilibrium systems, the observed transition is a many-body effect. This is evidenced by the required critical number of atoms, which is finite but large, typical of phase transitions in cold atom systems \cite{Anderson1995,Davis1995,Damle1996,Niu2006,Donner2007}. However, in contrast to systems in thermal equilibrium, our system is not characterized by the free energy and has a discrete time-translation symmetry. Most importantly, there are no long-range correlations of spatial density fluctuations. The intra-cloud density fluctuations are irrelevant, they are uncorrelated with the inter-cloud fluctuations, which, as we show, are responsible for the transition and display critical slowing down.  As a result, the symmetry breaking is quantitatively described by the mean-field theory.

The irrelevance of intra-cloud density fluctuations is a consequence of the time scale separation and the weakness of the interaction. The decay of correlations of intracloud density fluctuations is characterized by the damping rate $\Gamma$ of atomic motion, which is determined by the Doppler effect in the MOT \cite{Sesko1991} and does not change at the phase transition. On the other hand, switching between periodic states in modulated systems requires a large fluctuation and involves overcoming an effective activation barrier \cite{Chan2007,Dykman1998,Lapidus1999,Siddiqi2004,Vijay2009,Aldridge2005}. For low temperature, switching is a rare event. We measured the rates $W_{nm}$ of inter-cloud $n\rightarrow m$ switching ($n,m=1,2$) directly for low $N_{\rm tot}$ from the decay time of the difference of the cloud populations, which gave $W_{12}=W_{21}\sim 1$~s$^{-1}$. Therefore, in the experiment there is a strong inequality between the characteristic reciprocal times, $\omega_F\gg \Gamma \gg W_{nm}$.

During switching an atom is far from the clouds. Its interaction energy
with individual atoms in the clouds is much smaller than $k_BT$. However, the total energy of interaction with all atoms in the clouds may exceed $k_BT$, and then it changes the inter-cloud switching rate. The interaction depends on the density distribution within the clouds and ultimately on the cloud populations $N_{1,2}$.

The switching rates modification by the atom-atom interaction is the underlying mechanism of the symmetry breaking. Since these are the cloud populations $N_{1,2}$ that control the switching rates and that are changed as a result of switching, we choose their mean relative difference $\eta=\langle N_2-N_1\rangle/N_{\rm tot}$  as the order parameter. A natural control parameter is the reduced total number of atoms $\theta=\left(N_{\rm tot}-N_c\right)/N_c$.

\begin{figure}[h]
  \centering
  \includegraphics[scale=0.7]{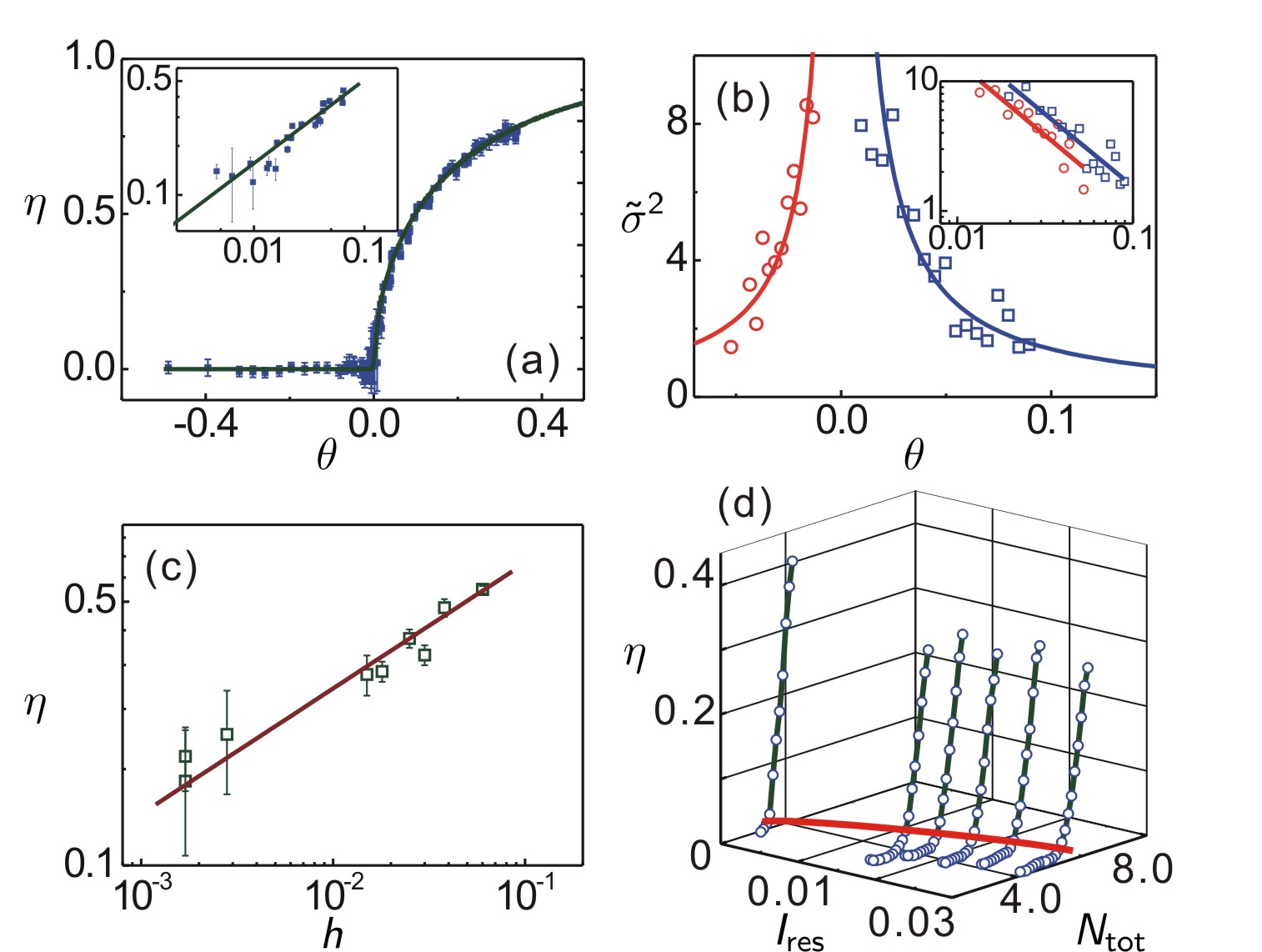}
  \caption{(Color) (a) The relative difference of the cloud populations $\eta=\langle N_2-N_1\rangle/N_{\rm tot}$ as a function of the control parameter $\theta=\left(N_{\rm tot}-N_c\right)/N_c$; the critical total number of atoms $N_c=1.6\times10^7$. The solid curves show the mean-field theory, $\eta=\tanh \left[(\theta+1)\eta\right]$. For $0<\theta\ll 1$ $|\eta|\propto\theta^{\beta}$ with $\beta=0.51\pm0.01$ (inset).  (b) The variance of the order parameter $\tilde\sigma^2=10^{3}\sigma^2$ and its scaling (inset). The critical exponents in the symmetric and broken-symmetry phases are, respectively, $\tilde{\gamma}=1.04\pm0.21$ and $\tilde{\gamma}=1.11\pm0.13$. (c) The order parameter at criticality ($\theta=0$) as a function of amplitude $h$ of the additional modulation at frequency $\omega_F/2$; $h$ is scaled by the strong modulation amplitude. The solid line is $|\eta|\propto h^{1/\delta}$, with $\delta=3.0\pm0.8$. (d) Order parameter $|\eta|$ as a function of $N_{\rm tot}$ (in units of $10^7$ atoms) and the intensity of the additional resonant radiation (scaled by the saturation intensity, 3.78 mW/cm$^2$). The red curve is the phase transition line. The error bars in panels (a) and (c) show the standard deviations of 10 measurements. }\label{fig2}
\end{figure}

\begin{figure}[h]
  \centering
  \includegraphics[scale=0.7]{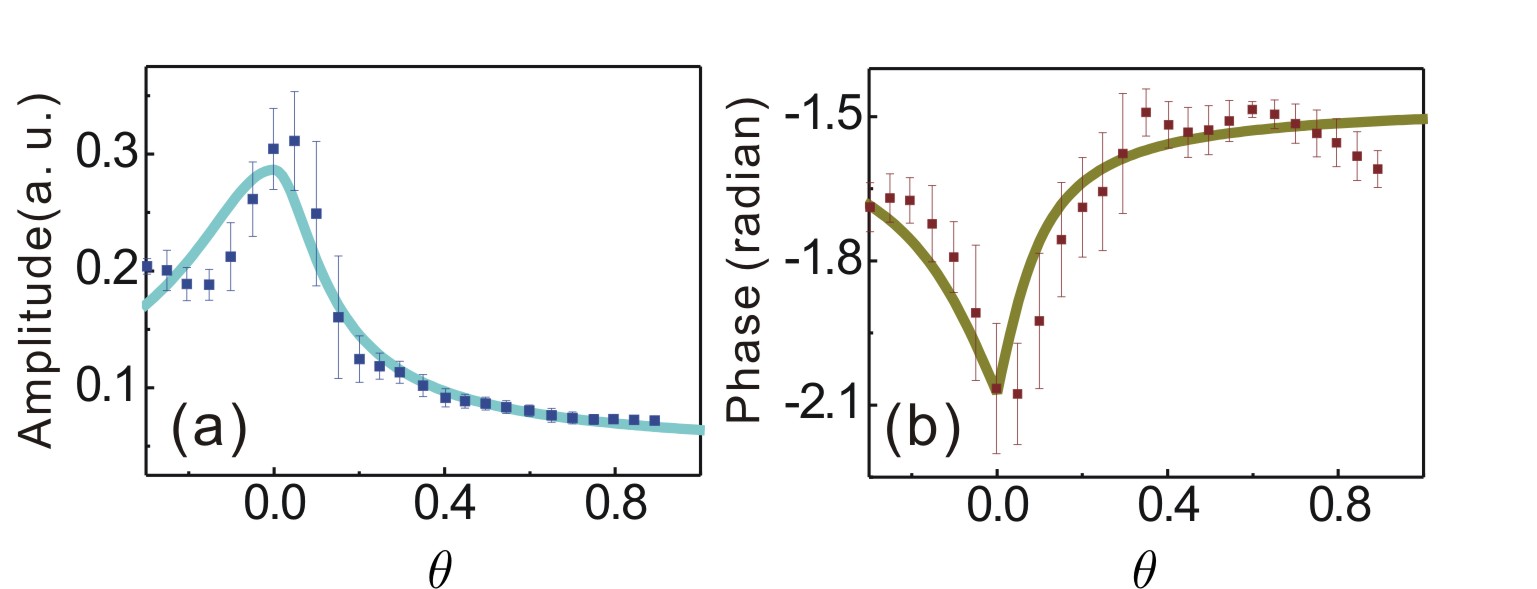}
  \caption{(Color online) The amplitude (a) and phase (b) of oscillations of the order parameter $\eta$ induced by an extra modulation at frequency $\omega_F/2+\Omega$ with $\Omega$=0.1 Hz. The solid curves show the theory with the $N_{\rm tot}\rightarrow 0$ switching rate used as a fitting parameter. The error bars show the standard deviations of 100 measurements.  }\label{fig3}
\end{figure}

Our main observations are presented in Figs.~\ref{fig2} and \ref{fig3}. They show that all measured quantities display mean-field critical behavior. The order parameter scales as  $|\eta|\propto\theta^{\beta}$ for $\theta>0$, with $\beta\approx 1/2$, see Fig.~\ref{fig2}a, and $\eta=0$ for $\theta<0$ (this scaling was also seen earlier \cite{Kim2006}, but now we have come much closer to the critical point, and the precision is much higher). Fluctuations of the population difference are characterized by the variance $\sigma^2=N_{\rm tot}^{-2}\langle(N_2-N_1)^2\rangle-\eta^2$. It scales as $\sigma\propto|\theta|^{-\tilde\gamma}$ with $\tilde\gamma\approx1$ for positive and negative $\theta$, see Fig.~\ref{fig2}b. Equilibrium systems at criticality show a strongly nonlinear response to the symmetry-breaking static field, like a magnetic field at the Ising transition. For our system an analogue of such field is an additional periodic driving at half the strong-field frequency. With this field, the overall time-translation period is $2\tau_F$, and the vibrational states of period $2\tau_F$ become non-equivalent \cite{Ryvkine2006a}. We studied the effect of the dynamic symmetry breaking field at criticality, $\theta=0$, by asymmetrically modulating the counter propagating beams. Fig.~\ref{fig2}c shows that the cloud population difference scales with the additional field amplitude $h$ as $|\eta|\propto h^{1/\delta}$ with $\delta\approx3$, in agreement with the mean field theory.

The critical slowing down should lead to a strong response of the population difference to a field detuned by a small frequency $\Omega$ from $\omega_F/2$. This is an analog of applying a weak slowly varying field to thermal equilibrium systems. The amplitude and phase of the forced oscillations of $\eta$ at frequency $\Omega$ as functions of $\theta$ are shown in Fig.~\ref{fig3}. They display characteristic asymmetric resonant features, again in excellent agreement with the mean-field theory.

In general, one should not expect that our oscillating system can be described by the Landau-type theory with standard critical exponents, and in fact the theory has to be extended. We explain the observations as due to kinetic many-body effects. They turn out to be strong as a consequence of the exponential sensitivity of the inter-cloud switching rates $W_{nm}$.

If the atom-atom coupling can be disregarded, the switching rates are equal by symmetry, $W_{12}^{(0)}=W_{21}^{(0)}\equiv W^{(0)}=C\exp \left(-R^{(0)}/k_BT\right)$ with $C\sim\omega_{\rm cl}$, where $\omega_{\rm cl}\approx 3\Gamma$ is the frequency of small-amplitude damped atom vibrations about the cloud center. Of primary interest is the activation energy $R^{(0)}$. For weekly non-sinusoidal period-2 vibrations it was calculated and measured for single-oscillator systems with cubic nonlinearity \cite{Chan2007,Dykman1998,Lapidus1999}. If we use the same model, for the present experimental parameters the theory \cite{Dykman1998} gives $R^{(0)}/k_B\approx$3.4~mK. This is within a factor of 2 of the value extracted from the measured $W^{(0)}$ and estimated $T$, which is satisfactory given that in the present case the parametric modulation was comparatively strong (the squared MOT eigenfrequency $\omega_0^2$ was modulated with relative amplitude 0.9), and therefore the vibrations were noticeably non-sinusoidal.

The atom-atom interaction changes the switching activation energy. The change becomes significant only because of the cumulative effect of the interaction of a switching atom with the atoms in the clouds. For not too strong interaction the rate of $1\rightarrow 2$ switching becomes
\begin{equation}
\label{eq:ratesmany_atom}
W_{12}(N_1;N_{\rm tot})=W^{(0)}\exp[\alpha (N_2-N_1)+\beta N_{\rm
tot}]
\end{equation}
with $\alpha,\beta\propto 1/T$ (to obtain $W_{12}$ one should replace $ N_1\leftrightarrow N_2$). The parameters $\alpha, \beta$ are determined by the generalized work of the interatomic force done on an atom during its intercloud switching, see Supplementary Material. The probability $P_1$ to have $N_1$ atoms in cloud 1 for given $N_{\rm tot}$ can be found from the balance equation. For $N_{\rm tot}\gg 1$, the stationary probability $P_1^{\rm st}(N_1)$ is a function of the reduced population difference $x=(N_2-N_1)/N_{\rm tot}$ and the control parameter $\theta =\alpha N_{\rm tot}-1$. It has a sharp maximum whose position gives the order parameter $\eta=\langle x\rangle$.

For fixed $N_{\rm tot}$, near criticality ($|x|,|\theta|\ll 1$) the stationary distribution has a characteristic Landau mean-field form $P_1^{\rm st}\propto\exp[-N_{\rm tot}(x^4-6\theta x^2)/12]$ \cite{LL_statphys1}. For $\theta<0$ it has a peak at $x=\eta=0$, which corresponds to equal mean cloud populations. For $\theta>0$ and $|\theta|\gg N_{\rm tot}^{-1/2}$ the symmetry is broken and $\langle x\rangle =\eta=\pm(3\theta)^{1/2}$. This describes the dependence of $\eta$ on $\theta$ in Fig.~\ref{fig2}a. The form of $P_1^{\rm st}$ explains also the scaling $\sigma^2\propto |\theta|^{-1}$ of the variance $\sigma^2=\langle x^2\rangle-\langle x \rangle^2$ in Fig.~\ref{fig2}b.

The critical total number of atoms (where $\theta=0$) is $N_c=\alpha^{-1}$. Thus, $N_c \propto T$. This dependence was tested by heating up the atoms with additional resonant radiation. Indeed, $N_c$ was found to increase almost linearly with the radiation intensity, see Fig.~\ref{fig2}d.

An extra modulation of the beam intensities $h \cos(\omega_F t/2+\phi_h)$
leads to factors $\exp(h_{12})$ and $\exp(-h_{12})$ in the switching rates $W_{12}$ and $W_{21}$, with $h_{12} \propto h/T$ \cite{Ryvkine2006a}. In calculating these factors one can disregard the weak atom-atom interaction. As a result, $P_1^{\rm st}$ is multiplied by $\exp(N_{\rm tot}h_{12}x)$. This gives $\eta \propto h^{1/3}$ at criticality ($\theta=0$), in agreement with the experiment. If the frequency of the extra modulation is slightly detuned from $\omega_F/2$, the linear response does not diverge at criticality. The result of the mean-field calculation in the Supplementary Material is in excellent agreement with the experiment, as seen from Fig.~\ref{fig3}.

An independent estimate of the interaction strength can be obtained from the dependence of the vibration amplitudes of the clouds on the cloud populations. For comparatively weak interaction, the interaction-induced change of the amplitude of an $n$th cloud should be proportional to the number of atoms in the other cloud $N_{3-n}$ ($n=1,2$). This is indeed seen in Fig.~\ref{fig1}c. The main contribution comes from the long-range interaction, the shadow effect \cite{Dalibard1988,Sesko1991,Mendoncca2008}. This interaction also gives the main contribution to the parameters $\alpha, \beta$ and thus determines the critical total number of atoms $N_c$. The value of $N_c$ obtained in a simple one-dimensional model \cite{Dykman1998,Ryvkine2006a} that assumes sinusoidal vibrations is within a factor of 2 from the experimental data, as are also the slopes of the straight lines in Fig.~\ref{fig1}c.

In the experiment, the total number of trapped atoms was slowly fluctuating. The mean-field theory still remains applicable, but needs to be extended, see Supplementary Material. An important consequence is that in the low-symmetry phase, $1\gg \theta\gg N_c^{-1/2}$, the variance $\sigma^2$ is modified, $N_c\sigma^2=(2\theta)^{-1}+ 3(4\theta+\ep)^{-1}$, where $\ep=W_{\rm out}/W^{(0)}\exp(\beta N_c)$, $\ep \ll 1$, with $W_{\rm out}$ being the probability for an atom to leave the trap per unit time. The variance is larger than in the symmetric phase for the same $|\theta|$ by factor 5/4 for $\theta \gg \ep$ and is smaller by factor 2 for smaller $\theta$, i.e., closer to $N_c$. The scaling $\sigma^2 \propto \theta^{-1}$ holds in the both limits. This is the scaling seen in Fig.~\ref{fig2}b; we could not come close enough to $N_c$ to observe the crossover; $\sigma^2$ in the broken-symmetry phase remained larger than in the symmetric phase.

In conclusion, we demonstrate the occurrence of an ideal mean-field transition far from equilibrium,  the spontaneous breaking of the discrete time-translation symmetry in a system of periodically modulated trapped atoms. The mean-field behavior is evidenced by the critical exponents of the order parameter and its variance, as well as the nonlinear resonant response at criticality and the susceptibility as a function of the distance to the critical point. We explain the effect as resulting from the interplay of the interaction and nonequilibrium fluctuations, where the interaction, even though comparatively weak, is strong enough to affect the rates of fluctuation-induced transitions between coexisting vibrational states. The proposed theory is in full agreement with the observations.

We thank M. C. Cross, G. S. Jeon, W. D. Phillips, D. L. Stein, and T. G. Walker for helpful comments. This work was supported by the Acceleration Research Program of Korean Ministry of Science and Technology. MD was supported in part by NSF grant PHY-0555346. WJ acknowledges support from Yeonam Foundation.

%\bibliographystyle{apsrev}
%\bibliography{C:/Aaa/BibTex/md10,MyBibliography1}

\end{document}

\title{Supplementary Material for ``An Ideal Mean-Field Transition in a Modulated Cold Atom System"}
\author{Myoung-Sun Heo}
\affiliation{%
Department of Physics and Astronomy, Seoul National University, Seoul 151-747, Korea
}%
\author{Yonghee Kim}
\affiliation{%
Department of Physics and Astronomy, Seoul National University, Seoul 151-747, Korea
}%
\author{Kihwan Kim}
\affiliation{%
Department of Physics and Astronomy, Seoul National University, Seoul 151-747, Korea
}%
\author{Geol Moon}
\affiliation{%
Department of Physics and Astronomy, Seoul National University, Seoul 151-747, Korea
}%
\author{Junhyun Lee}
\affiliation{%
Department of Physics and Astronomy, Seoul National University, Seoul 151-747, Korea
}%
\author{Heung-Ryoul Noh}
\affiliation{%
Department of Physics, Chonnam National University, Gwangju 500-757, Korea
}%
\author{M. I. Dykman}
\affiliation{%
Department of Physics and Astronomy, Michigan State University, East Lansing, Michigan 48824, USA
}%
\author{Wonho Jhe}
\affiliation{%
Department of Physics and Astronomy, Seoul National University, Seoul 151-747, Korea}

\date{\today}

\maketitle
\section{Theory of inter-cloud switching}
\subsection{Single-atom switching}

Here we provide a detailed closed-form description of the critical phenomena in a parametrically modulated MOT. We start with the single-atom dynamics for resonant MOT modulation, where the modulation frequency $\omega_F$ is close to twice the eigenfrequency $\omega_0$ of atomic vibrations in the MOT. The modulation can lead to the onset of period-two vibrations at frequency $\omega_F/2$. These vibrations are close to sinusoidal for moderately strong modulation, there is no dynamical chaos in the range of phase space of interest. In the single-atom approximation the equation of motion can be written as
\begin{eqnarray}
\label{eq:eom_oneatom_general}
m_a\ddot \rb = {\bf G}(\rb,\dot\rb;t)+\fb(t)
\end{eqnarray}
where $\rb$ is the position vector of an atom, $m_a$ is the atomic mass, and ${\bf G}$ is the well-understood force from the laser beams \cite{Sesko1991,Steane1992,Arnold2000}; for modulated beams it periodically depends on time, ${\bf G}(t+\tau_F)={\bf G}(t)$, with $\tau_F=2\pi/\omega_F$ being the modulation period. The force $\fb(t)$ in Eq.~(\ref{eq:eom_oneatom_general}) is the zero-mean noise from spontaneous light emission, which is often modeled by white Gaussian noise, $\langle f_{\kappa}(t)f_{\kappa'}(t')\rangle = 4\Gamma m_a k_BT\delta_{\kappa\kappa'}\delta(t-t')$, where $\Gamma$ is the effective friction coefficient, $T$ is the characteristic gas temperature, and $\kappa,\kappa' =1,2,3$ enumerate the atomic coordinates.

We are interested in the parameter range where, in the absence of noise, Eq.~(\ref{eq:eom_oneatom_general}) has two stable periodic solutions $\rb_{1,2}(t)$, with $\rb_n(t+2\tau_F)=\rb_n(t)$ ($n=1,2$) and $\rb_2(t)=\rb_1(t+\tau_F)$. These solutions give the positions of the centers of the vibrating atomic clouds in Fig. 1a of the main paper. By linearizing Eq.~(\ref{eq:eom_oneatom_general}) about $\rb_n(t)$ one finds the typical frequencies $\omega_{\rm cl}$ and relaxation rates ($\sim\Gamma$) of the intracloud motion.  Noise leads to fluctuations about $\rb_n(t)$ with variance $\propto T$ that determines the size of the clouds in Fig.~1.

Switching between the clouds results from large rare fluctuations. It is central for the analysis that the switching rates are much smaller than $\omega_F$ and than $\omega_{\rm cl},\Gamma$. Therefore of interest are period-averaged rates $W_{nm}$ of $n\to m$ switching, $n,m=1,2$. Quite generally \cite{Freidlin_book,Dykman2001}
\begin{eqnarray}
\label{eq:Wnm_general}
W_{nm}=C_W\exp(-R_n/k_BT),
\end{eqnarray}
where the prefactor $C_W\sim\max(\omega_{\rm cl},\Gamma)$. The most important in $W_{nm}$ is the exponential factor, which has the activation form. The activation energy $R_n$ of switching from the periodic state $n$ is given by the solution of a variational problem $R_n=\min{\cal R}$, where the functional ${\cal R}$ is \cite{Dykman2001},
\begin{eqnarray}
\label{eq:var_probl_R}
{\cal R}=(8\Gamma m_a)^{-1}\int dt \fb(t)^2 +\int dt{\bm\lambda}(t)\left[m\ddot\rb - {\bf G}-\fb(t)\right]
\end{eqnarray}
Here, the term quadratic in $\fb$ determines the probability density of a realization of the random force that leads to switching \cite{FeynmanQM}; ${\bm\lambda}(t)$ is the Lagrange multiplier that relates the dynamics of the atom and the force to each other. For switching from $n$th periodic state, the trajectory $\rb(t)$ should go from $\rb_n(t)$ for $t\to -\infty$ to the unstable periodic saddle-type state $\rb_b(t)$ on the boundary of  the basin of attraction to $\rb_n(t)$ for $t\to\infty$; ${\bm\lambda}(t), \fb(t)\to {\bf 0}$ for $t\to\pm\infty$.

Since in the single-atom case the period-two vibrational states differ only by phase, the activation energies $R_{1,2}$ are equal, $R_1=R_2=R^{(0)}$. The variational problem is significantly simplified if the  vibrations $\rb_n(t)$ are almost sinusoidal, which corresponds to small damping and comparatively weak nonlinearity, see below. For a one-dimensional weakly nonlinear oscillator $R^{(0)}$ was found and measured earlier \cite{Dykman1998,Lapidus1999,Chan2007}.

\noindent\subsection{Switching rate modification by the atom-atom interaction}

The interaction between atoms leads to an extra force ${\bf G}_{\rm int}(\rb)$ in Eq.~(\ref{eq:eom_oneatom_general}). This force is weak in the sense that it weakly affects the intracloud atomic dynamics. In the experiment, the radius of the clouds changed by $\sim 10\%$ where the number of atoms changed by a factor of 2. The change of the vibration amplitude was also small, see Fig.~1 of the main paper.

The force ${\bf G}_{\rm int}$ depends on the positions of all atoms. In the analysis of the effect of the interaction on the intracloud dynamics it can be separated into the mean-field and fluctuating parts. The mean-field part is due to the long-range interaction and is described in terms of the continuous atomic density \cite{Sesko1991}. The fluctuating part is dominated by short-range elastic collisions. For comparatively low atomic densities and high temperatures used in the experiment the random part of  ${\bf G}_{\rm int}$ is small compared to the thermal noise and will be disregarded. The inter-cloud interaction is determined by long-range forces and can be described in the mean-field approximation: the force ${\bf G}_{\rm int}(\rb)$ from another cloud changes only slightly when atoms in this cloud shift by an interatomic distance.

From the above arguments, the interatomic force on an atom in cloud $n$,  ${\bf G}_{\rm int}(\rb;n)$, can be written  as
\begin{eqnarray}
\label{eq: interatomic_force_ito_density}
{\bf G}_{\rm int}(\rb;n) =\sum\nolimits_{m=1,2} N_m{\bf G}_{\rm int}(\rb;n|m),
\end{eqnarray}
where $N_{m}$ is the number of atoms in cloud $m$. The force ${\bf G}_{\rm int}(\rb;n|m)$ is determined by the density distribution in the cloud $m$, and the terms with $n=m$ and $n\neq m$ describe the force from the atoms in the same and in the different cloud, respectively.

Even though ${\bf G}_{\rm int}$ is small compared to the single-atom force ${\bf G}$, it may strongly affect the switching rate, leading to the spontaneous symmetry breaking. This is because ${\bf G}_{\rm int}$ changes the  switching activation energy $R_n$. Extending to the many-atom system the approach of Ref.~\onlinecite{Dykman2001}, to first order in the interaction we obtain $R_n=R^{(0)}+R_n^{(1)}$ with
\begin{eqnarray}
\label{eq:many_atom_correct_to_Rn}
&&R_n^{(1)}=\sum_m\alpha_{nm}N_m,\nonumber\\
&&\alpha_{nm}=-\int_{-\infty}^{\infty}dt {\bm\lambda}_n^{\rm (opt)}(t){\bf G}_{\rm int}\bigl(\rb_{n}^{\rm (opt)}(t);n|m\bigr).
\end{eqnarray}
Here, $\rb_{n}^{\rm (opt)}(t)$ and ${\bm\lambda}_n^{\rm (opt)}(t)$ describe the trajectory that minimizes functional ${\cal R}$, Eq.~(\ref{eq:var_probl_R}), for switching from state $n$. Function $\rb_{n}^{\rm (opt)}(t)$ gives the most probable path followed in switching; for a parametrically excited oscillator this path was recently observed in experiment \cite{Chan2008}. Since the states $1$ and $2$ differ only in phase, we have $\alpha_{11}=\alpha_{22}$ and $\alpha_{12}=\alpha_{21}$.

The following arguments were used in deriving Eq.~(\ref{eq:many_atom_correct_to_Rn}). First, because the interatomic interaction is weak, atoms switch one by one, not in groups; switchings of individual atoms are uncorrelated. Second, and most importantly, in the optimal fluctuation leading to switching of an atom the atomic clouds practically do not change their shape. This is also because the interatomic interaction is weak, and the force from a single atom in the cloud very weakly affects the motion of the switching atom, $|\alpha_{nm}|\ll k_BT$. Only the effect from all atoms in the cloud is appreciable, the product $\alpha_{nm}N_m$ ($N_{1,2}\gg 1$) can become larger than $k_BT$. On the other hand, a significant change of the shapes of the clouds, even though it would change the switching activation energy, has an exceedingly small probability, which can not compensate the decrease of $R_{1,2}^{(1)}$. Indeed, if the cloud radius is $a_0\sim (k_BT/m_a\omega_{\rm cl}^2)^{1/2}$, the probability density to have the atoms in $n$th cloud displaced by $\Delta r_n\gg a_0$ is $\propto \exp(-N_n \Delta r_n^2/2a_0^2)$. Therefore in calculating ${\bf G}_{\rm int}\bigl(\rb_{n}^{\rm (opt)}(t);n|m\bigr)$ one should assume that the atoms in clouds $1$ and $2$ are at their stable equilibrium positions $\rb_1(t)$ and ${\bf r}_2(t)$. In other words, there is no fluctuational barrier preparation \cite{Kagan1976,Kagan1992} for intercloud switching, even though the switching barrier is strongly affected by the many-body interaction. This significantly simplifies the explicit calculation.

Equation (\ref{eq:many_atom_correct_to_Rn}) shows that the effective switching activation energy linearly depends on the numbers of atoms in the clouds, for weak interatomic coupling. Even though $|R_n^{(1)}|$ is small compared to the single-atom activation energy $R^{(0)}$, it can largely exceed $k_BT$, in which case the change of the switching rate will be large. From Eqs.~(\ref{eq:Wnm_general}) and (\ref{eq:many_atom_correct_to_Rn}) we obtain for the many-atom switching rate
\begin{eqnarray}
\label{eq:switching_rate_with_interaction}
&W_{nm}\equiv W_{nm}(N_n;N_{\rm tot})=W^{(0)}e^{(\alpha+\beta)N_{\rm tot}-2\alpha N_n}, \nonumber\\
&N_{\rm tot}=N_1+N_2,
\end{eqnarray}
where $W^{(0)}=C_W\exp(-R^{(0)}/T)$ is the switching rate in the neglect of the interatomic interaction; $\alpha=(\alpha_{11}-\alpha_{12})/2k_BT$ and $\beta=-(\alpha_{11}+\alpha_{12})/2k_BT$. Equation (\ref{eq:switching_rate_with_interaction}) coincides with Eq.~(1) of the paper and provides a general form of the coefficients $\alpha,\beta$.

\section{The symmetry breaking transition}

\subsection{ The master equation}

Because atoms move fast within the clouds compared to inter-cloud switching, on time scale $\sim W_{nm}^{-1}$ the state of the system is {\it fully characterized} by the numbers of atoms in each cloud $N_1,N_2$. For a given $N_{\rm tot}$, the system is described by probability $P_1(N_1)$ to have $N_1$ atoms in cloud 1 at time $t$. This probability can be found from the master equation
\begin{widetext}
\begin{eqnarray}
\label{eq:master_eqtn}
&&\dot P_1(N_1)=-[\mu(N_1)+\nu(N_1)]P_1(N_1)+\nu(N_1-1)P_1(N_1-1) + \mu(N_1+1)P_1(N_1+1),\nonumber\\
&&\mu(N_1)=N_1W_{12}(N_1;N_{\rm tot}), \quad\nu(N_1)=(N_{\rm tot}-N_1)W_{12}(N_{\rm tot}-N_1;N_{\rm tot}).
\end{eqnarray}
\end{widetext}
Here we have taken into account that, as a result of a  $n\to m$ transition, the number of atoms in cloud $n$ decreases by $1$ and in cloud $m$ increases by $1$, and that the total probability of such a transition per unit time for $N_n$ atoms in cloud $n$ is $N_nW_{nm}$.

Prior to making a transition the atom will move randomly within the cloud for a long time and will completely ``forget'' its initial state. It is this randomization that makes the many-particle system zero-dimensional and leads to the mean-field approximation being exact.

The stationary solution of Eq.~(\ref{eq:master_eqtn}) is
\begin{eqnarray}
\label{eq:stationary_dist_no_modulation}
P_1^{\rm st}(N_1)= Z^{-1}\binom{N_{\rm tot}}{N_1}\exp\left[-2\alpha N_1(N_{\rm tot}-N_1)\right],
\end{eqnarray}
where $Z\equiv Z(N_{\rm tot})$ is the normalization constant given by condition $\sum\nolimits_{N_1}P_1(N_1)=1$.

For large $N_{\rm tot}$, the function $P_1^{\rm st}(N_1)$ has one or two sharp peaks. The location of the peak(s) depends on the total number of atoms. We will study the critical region where $N_{\rm tot}$ is close to $1/\alpha$, in which case near the peak(s) $|N_1-N_2|\ll N_{\rm tot}$.   It is convenient then to introduce a quasi-continuous variable $x=(N_2-N_1)/N_{\rm tot}=1-2N_1/N_{\rm tot}$. For $|x|\ll 1$ from Eq.~(\ref{eq:stationary_dist_no_modulation}) we find
\begin{eqnarray}
\label{eq:distrib_small_x}
&P_1^{\rm st}(N_1)\approx \tilde Z^{-1}\exp\left[-N_{\rm tot}(x^4-6\theta x^2)/12\right],\nonumber\\
&\theta=\alpha N_{\rm tot}-1,
\end{eqnarray}
where $\tilde Z$ is the normalization constant.

Equation (\ref{eq:distrib_small_x}) has the standard form of the mean-field probability distribution near a symmetry-breaking transition \cite{LL_statphys1}.  The parameter $\theta$ is the control parameter, which plays the role of the deviation from the critical temperature in systems in thermal equilibrium, whereas $x$ plays the role of the order parameter. Parameter $\theta$ linearly depends on the total number of atoms; it also linearly depends on the reciprocal temperature. The critical number of atoms $N_c$ is determined by condition $\theta =0$,
\begin{equation}
\label{eq_N_c_define}
N_c=1/\alpha \propto T.
\end{equation}

\noindent\subsection{Critical exponents}

The mean reduced difference of the cloud populations $\eta=\langle x\rangle $ is determined by the position $x_0$ of the maximum of the distribution $P_1^{\rm st}(N_1)$. For $\theta < 0$ the distribution has one maximum at $x_0=\eta=0$, whereas at $\theta>0$ it has two symmetric maxima. The system occupies one of them, which corresponds to the spontaneous symmetry breaking. Close to the critical point
\begin{eqnarray}
\label{eq:critical_order_parameter}
\eta = N_{\rm tot}^{-1}\langle N_2-N_1\rangle  = \pm (3\theta)^{1/2}\quad {\rm for } \; 0<\theta \ll 1,
\end{eqnarray}
which is the familiar mean-field scaling with the critical exponent $1/2$.

Close, but not too close to the critical point, $1\gg |\theta|\gg N_c^{-1/2} $, from Eq.~(\ref{eq:distrib_small_x}) we obtain for the mean square fluctuations of the populations difference $\sigma^2=N_{\rm tot}^{-2}\langle (N_2-N_1)^2\rangle-\eta^2$
\begin{eqnarray}
\label{eq:population_fluct_symmetric_no_field}
&\sigma^2 = (N_c|\theta|)^{-1}\quad {\rm for}\; \theta<0; \nonumber\\
&\sigma^2=(2N_c|\theta|)^{-1}\quad {\rm for}\; \theta> 0
\end{eqnarray}
($|\theta|\ll 1$). This shows the familiar $\theta^{-1}$ scaling of the variance of the order parameter on the both sides of the critical point. At the critical point, $\theta=0$, we have $\sigma^2
 = 2\pi\sqrt{6}N_c^{-1/2}/\Gamma(1/4)^2\approx 1.2N_c^{-1/2}.$
Note that $\langle x^2\rangle$ remains finite in a finite-size system at the critical point, but its dependence on $N_{\rm tot}\approx N_c$ is given by factor $N_{\rm tot}^{-1/2}$ instead of $N_{\rm tot}^{-1}$ far from criticality.

\subsection{Response to the symmetry-breaking field}

The symmetry of the period-two vibrational states can be lifted if the system is driven by an extra additive force ${\bf h}(t)={\bf h}\cos(\omega_Ft/2+\phi_h)$. Such force is an analog of a static force in a thermal equilibrium system, it is ``static" in the frame oscillating at frequency $\omega_F/2$. In the experiment the force was produced by modulating the counter-propagating laser beams at frequency $\omega_F/2$ with different amplitude; the phase $\phi_h$ is counted off from the phase of the strong modulation with period $\tau_F$. Response to a period-$2\tau_F$ force becomes strongly nonlinear at the critical point $\theta=0$ even for small amplitude $|{\bf h}|$, because the force modifies the switching activation energies $R_{1,2}$ \cite{Ryvkine2006a}. The latter can be found from Eq.~(\ref{eq:var_probl_R}) with ${\bf G}\to {\bf G}+{\bf h}(t)$. Extending the analysis of Refs.~\onlinecite{Dykman2001,Ryvkine2006a}, we obtain to first order in the inter-atomic interaction and the symmetry-lifting field
\begin{eqnarray}
\label{eq:activ_energy_with_lifted_symmetry}
&&R_n=R^{(0)} + R_n^{(1)}+ R_n^{(h)}, \nonumber\\
&&R_n^{(h)}=-\int dt {\bm \lambda}_n^{\rm (opt)}(t){\bf h}(t)=\bar h \cos \phi_n.
\end{eqnarray}
The terms $R_{1,2}^{(h)}$ are proportional to the field amplitude, $\bar h\propto |{\bf h}|$. The phases $\phi_{1,2}$ are the same as for a single parametrically modulated oscillator additionally driven by a force at frequency $\omega_F/2$ \cite{Ryvkine2006a}. They are linear in $\phi_h$, with $\phi_2=\phi_1+\pi$. This can be immediately seen from Eq.~(\ref{eq:activ_energy_with_lifted_symmetry}), since by symmetry ${\bm \lambda}_2^{\rm (opt)}(t)= \lab_1^{\rm (opt)}(t+\tau_F)$, whereas ${\bf h}(t+\tau_F)=-{\bf h}(t)$. As a consequence, $R_1^{(h)}=-R_2^{(h)}$. The difference in the activation energies of $1\to 2$ and $2\to 1$ transitions lifts the degeneracy of the states 1 and 2, making their populations different even in the absence of the inter-atomic coupling.

Even though $|R_n^{(h)}|\ll R^{(0)}$ for small field amplitude $|{\bf h}|$, the ratio $R_n^{(h)}/k_BT$ should not be small, and therefore the field-induced change of the switching rates may be substantial. At the same time, the change of the amplitudes of cloud vibrations remains small for small ${\bf h}$ and will not be discussed. Equation (\ref{eq:activ_energy_with_lifted_symmetry}) for $R_n^{(h)}$ resembles the change of the free energy of an Ising spin by a magnetic field $\propto \bar h$ tilted by an angle $\phi_1$ with respect to the quantization axis.

Formally, the terms $R_{1,2}^{(h)}$ lead, respectively, to factors $\exp[-(\bar h/k_BT)\cos\phi_1]$  and $\exp[(\bar h/k_BT)\cos\phi_1]$ in the rate coefficients $\mu(N_1)$ and $\nu(N_1)$ in master equation (\ref{eq:master_eqtn}). As a result the stationary distribution $P_1^{\rm st}(N_1)$, Eq.~(\ref{eq:distrib_small_x}), acquires an extra factor
\begin{eqnarray}
\label{eq:field_modified_G}
P_1^{\rm st}(N_1)\to P_1^{\rm st}(N_1)\exp\left[-(N_{\rm tot}\bar h x/k_BT)\cos\phi_1\right]
\end{eqnarray}
[the normalization constant in $P_1^{\rm st}$ is changed appropriately; we remind that $x=(N_2-N_1)/N_{\rm tot}$.].

The equation for the maximum of the distribution $x_0=\eta$ near criticality now becomes
%
%\begin{eqnarray}
%\label{eq:eqn_for_order_par_with_modulation}
$\theta x_0-\frac{1}{3}x_0^3-(\bar h/k_BT)\cos\phi_1=0.$
%\end{eqnarray}
%
For small $|\bar h/k_BT|\ll 1$ and not too close to the phase-transition point, $1\gg|\theta|\gg N_c^{-1/2}$, we have
\begin{eqnarray}
\label{eq:linear in_f__order_parametr}
&\eta\approx \theta^{-1}(\bar h/k_BT)\cos\phi_1\quad (\theta<0),\\
&\eta\approx \pm (3\theta)^{1/2}- (2\theta)^{-1}(\bar h/k_BT)\cos\phi_1 \quad (\theta > 0).\nonumber
\end{eqnarray}
The generalized resonant susceptibility $\Xi(0)=-d\eta/d\bar h$ scales as $|\theta|^{-1}$ on the both sides of the phase transition point and diverges for $\theta\to 0$. We note that, even though we consider a system far from equilibrium and the susceptibility is calculated with respect to a time-dependent field, for $N_{\rm tot}=$~const it is simply related to the variance of the order parameter $\sigma^2$ as given by Eq.~(\ref{eq:population_fluct_symmetric_no_field}).

At the phase transition point, $\theta=0$, there is no linear static susceptibility, and $\eta =x_0 \propto \bar h^{1/3}$, as seen in the experiment.

\subsection{Frequency dispersion of resonant response}

The above analysis can be extended also to the case of an extra additive force with frequency $(\omega_F/2)+\Omega$ that differs from, but remains close to $\omega_F/2$, i.e., $|\Omega|\ll \Gamma\ll \omega_F$. To find the distribution $P_1$ near the maximum, one can transform Eq.~(\ref{eq:master_eqtn}) into a Fokker-Planck equation by expanding $P_1(N_1\pm 1)-P_1(N_1)\approx \pm \partial_{N_1}P_1+(1/2)\partial^2_{N_1}P_1$ and using similar expansions for $\mu, \nu$ \cite{vanKampen_book}. In the absence of extra modulation, for small $x=(N_2-N_1)/N_{\rm tot}$ to leading order in $x$ the equation reads
\begin{eqnarray}
\label{eq:FPE}
&\tilde W^{-1}\dot P_1=\hat L P_1,\qquad \tilde W=W^{(0)}\exp(\beta N_{\rm tot}),\nonumber\\
&\hat L P_1 =-2\parx\left[\bigl(\theta x-\frac{1}{3}x^3\bigr)P_1\right] + 2N_{\rm tot}^{-1}\parx^2P_1.
\end{eqnarray}
The stationary solution of Eq.~(\ref{eq:FPE}) $P_1^{\rm st}$ has the form (\ref{eq:distrib_small_x}).

An extra additive force ${\bf h}\cos\left[\left(\frac{1}{2}\omega_F+\Omega\right)t\right]$ with small $\Omega$ adiabatically modulates the rates $W_{nm}$ of the inter-cloud transitions, i.e., one can think of the rates $W_{nm}$ as parametrically dependent on time. The major contribution to the rate modulation comes from the modulation of the activation energies, which is described by Eq.~(\ref{eq:activ_energy_with_lifted_symmetry}) with
\begin{eqnarray}
\label{eq:periodic_modulation}
R_1^{(h)}\equiv R_1^{(h)}(t)=-R_2^{(h)}(t)= \bar h\cos(\Omega t)
\end{eqnarray}
(we have disregarded a time-independent term in the argument of the cosine; for quantities that do not oscillate at frequency $\omega_F/2$, like cloud populations, this term can be eliminated by changing time origin).

For weak modulation one can linearize the transition rates in $\bar h$ and consider the terms $\propto \bar h$ as a perturbation. Then the distribution can be sought in the form $P_1=P_1^{\rm st} + \delta P_1$, where the field-induced term $\delta P_1\propto \bar h$ is oscillating at frequency $\Omega$. It is given by equation
\begin{eqnarray}
\label{eq:eqn_linearized_in_x}
&&\tilde W^{-1}\delta\dot P_1 -\hat L\delta P_1 \nonumber\\
&&\qquad=2\frac{\bar h}{k_BT}\cos(\Omega t)N_{\rm tot}\left(\theta x - \frac{1}{3}x^3\right) P_1^{\rm st}.
\end{eqnarray}

Equation (\ref{eq:eqn_linearized_in_x}) allows one to find the periodically oscillating term in the order parameter $\delta\eta(t) = \langle x\rangle -x_0$ ($x_0$ is the position of the maximum of the stationary distribution) and the generalized susceptibility $\Xi(\Omega)$, which can be defined by expression
\[\delta\eta(t) = -\frac{1}{2k_BT}\left[\Xi(\Omega)\bar h\exp(-i\Omega t) +{\rm c.c.}\right].\]
This can be done by multiplying Eq.~(\ref{eq:eqn_linearized_in_x}) by $x$ and integrating over $x$.  Expanding $\theta x- x^3/3$  in  $\hat L$ about $x_0$, for $1\gg |\theta|\gg N_{\rm tot}^{-1/2}$ one finds
\begin{eqnarray}
\label{eq:susc_omega_explicit}
&&\Xi(\Omega)= 2\tilde W/(2|\theta|\tilde W -i\Omega), \qquad \theta < 0,\nonumber\\
&&\Xi(\Omega)= 2\tilde W/(4\theta\tilde W -i\Omega), \qquad \theta > 0.
\end{eqnarray}
[in the critical region, in Eq.~(\ref{eq:FPE}) for $\tilde W$ one should replace $N_{\rm tot}$ with $N_c$]. For $\Omega\neq 0$ the susceptibility Eq.~(\ref{eq:susc_omega_explicit}) remains finite even at the phase transition, $\theta =0$. Here $\Xi(\Omega)=2 i\tilde W/\Omega$. The susceptibility diverges with frequency, with critical exponent equal to $-1$.

\section{Fluctuations of the total trap population}

In the above analysis we disregarded inelastic collisions between trapped atoms and collisions with atoms outside the trap. These processes lead to fluctuations of the total number of atoms in the trap. In the experimental conditions they were slow, with rates an order of magnitude smaller than the rate of inter-cloud transitions $W_{nm}$. However, near the critical point inter-cloud population fluctuations are slowed down, and then fluctuations of the total number of trapped atoms $N_{\rm tot}$ may become important.

In the presence of fluctuations of $N_{\rm tot}$ the state of the system should be characterized by the probability $P({\bf N})$  [${\bf N}\equiv (N_1,N_2)$] to have $N_1$ atoms in cloud 1 and $N_2$ atoms in cloud 2, with $N_{\rm tot}=N_1+N_2$. For comparatively small $N_{\rm tot}$ and low temperatures used in the experiment, evaporation from the trap due to inelastic collisions between trapped atoms was rare. The major mechanism of the change of $N_{\rm tot}$ was collisions with atoms outside the trap. In a simple model of such collisions, the rate of loss of atoms from an $n$th cloud $W_{\rm out}N_n$ is proportional to the number of atoms in the cloud $N_n$, whereas the rate of capture of outside atoms by a cloud $W_{\rm out}N_0/2$ is independent of $N_n$ and depends only on the density of outside atoms (which determines $N_0$) and the cross-section of their capture. We disregard fluctuations in the density of outside atoms and choose parameter $N_0$ in such a way that it provides the typical scale of $N_{\rm tot}$, as will be seen below.

The master equation for $P({\bf N})$ has the form
\begin{widetext}
\begin{eqnarray}
\label{eq:master_w_escape}
 \dot P({\bf N}) &=&\tilde W\hat L_{\rm cl}P + W_{\rm out}\hat L_{\rm out}P, \qquad \hat L_{\rm cl}P= -\sum\nolimits_n\mu'(N_n;N_{\rm tot})P({\bf N})\nonumber\\
&& + \mu'(N_1+1;N_{\rm tot})P(N_1+1,N_2-1) + \mu'(N_2+1;N_{\rm tot})P(N_1-1,N_2+1),\\
\hat L_{\rm out}P&=&-(N_1+N_2+N_0)P({\bf N})+\sum\nolimits_n(N_n+1)P({\bf N}+{\bm\delta}_n)
%\nonumber\\
+(N_0/2)\sum\nolimits_nP({\bf N}-{\bm\delta}_{n}),\nonumber
\end{eqnarray}
\end{widetext}
where
\[\mu'(N_n;N_{\rm tot})=N_n\exp\left(\alpha N_{\rm tot}-2\alpha N_n\right)\equiv \tilde W^{-1}\mu(N_n)\]
and ${\bm \delta}_1=(1,0)$ and ${\bm \delta}_2=(0,1)$ [the inter-cloud switching rate $\tilde W$ is defined in Eqs.~(\ref{eq:switching_rate_with_interaction}), (\ref{eq:FPE})]. The terms $\hat L_{\rm cl}$ and $\hat L_{\rm out}$ describe inter-cloud transitions and exchange with atoms outside the trap, respectively. We assume that the typical rate of this exchange $W_{\rm out}\ll \tilde W$.

We now consider the critical region where $N_{\rm tot}$ is close to $N_c$. We will be interested in the range of $N_1,N_2$ where $P(N_1,N_2)$ is close to maximum, $|N_1-N_2|\ll N_{\rm tot}$, and introduce quasi-continuous variables
\begin{eqnarray}
\label{eq:x_and_u_variables}
&&x=\frac{N_2-N_1}{N_0}, \qquad u=\frac{N_1+N_2}{N_0}-1, \\
&&\theta = \alpha N_0-1.\nonumber
\end{eqnarray}
As we will see, variables $x$ and $\theta$ coincide with the variables used earlier if fluctuations of the total number of trapped atoms can be disregarded; variable $u$ gives the deviation of $N_{\rm tot}$ from $N_0$.

For $|x|\ll 1, |\theta|\ll 1$, and $|u|\ll 1$ to leading order in $x,\theta,u$, and $N_0^{-1}$ the operators $\hat L_{\rm cl}$ and $\hat L_{\rm out}$ become
%\begin{widetext}
%
\begin{eqnarray}
\label{eq:continuous_operators_intercloud_outside}
\hat L_{\rm cl}P&=& -2\parx\left\{\left[x(\theta +u)-\frac{1}{3}x^3\right]P\right\}\nonumber\\
&& + 2N_0^{-1}\parx^2P,\\
\hat L_{\rm out}P&=&  \parx(xP) + \partial_u(uP) +N_0^{-1}\left[(\parx^2+\partial_u^2)P\right].\nonumber
\end{eqnarray}
%\end{widetext}

Equations~(\ref{eq:master_w_escape}) - (\ref{eq:continuous_operators_intercloud_outside}) allow one to study the stationary distribution of the modulated system in the presence of fluctuations of the total population. We note first that exchange of atoms with the surrounding leads to effective exchange of atoms between the clouds. In turn, this leads to renormalization of inter-cloud transition rates and, effectively, of the control parameter,
\begin{eqnarray}
\label{eq:theta_renormalization}
\theta \to \theta -  \frac{W_{\rm out}}{2\tilde W},\qquad N_c\to \alpha^{-1}\left(1+\frac{W_{\rm out}}{2\tilde W}\right).
 \end{eqnarray}
The change of $\theta$ is negative, which should be expected, since exchange with the surrounding should stabilize the symmetric phase, and therefore a larger number of atoms is required for the symmetry breaking transition than in the case where there is no such exchange. $N_c$ is now the critical value of the parameter $N_0$, the mean number of trapped atoms is $\langle N_{\rm tot}\rangle = N_0(1+\langle u\rangle)$. The coefficient of diffusion over $x$ is also renormalized, $2N_0^{-1} \to 2N_0^{-1}(1+W_{\rm out}/2\tilde W)$.

In the stationary regime fluctuations of the total number of atoms in the trap are small, $\langle u^2\rangle \sim 1/N_0$. However, the distribution over the scaled difference of the cloud populations $x$ differs from the standard Landau-type distribution (\ref{eq:distrib_small_x}). This is in spite the fact that the system is described by the mean-field theory, no spatial correlations are involved. The dynamics of $x$ and $u$ are very different. There is no critical slowing down for $u$, and in the critical region $|\theta|\lesssim N_c^{-1/2}$ the width of the distribution over $x$ is $\sim N_c^{-1/4}\gg N_c^{-1/2}$.

\subsection{The variance of the order parameter}

The major effect of the coupling between fluctuations of $x$ and $u$ is the change of the variance $\sigma^2=\langle x^2\rangle-\langle x\rangle^2$. The latter occurs in the broken-symmetry state where $\theta\gg N_c^{-1/2}$. It can be calculated in the stationary regime by decoupling the chain of equations for the moments of the stationary distribution $\langle x^ku^l\rangle$ with integer $k,l$. For $N_c^{-1/2}\ll \theta\ll 1$
\begin{eqnarray}
\label{eq:variance_with_u_fluctuations}
N_c\sigma^2\approx \frac{1}{2  \theta}+\frac{3}{4  \theta + (W_{\rm out}/\tilde W)}.
\end{eqnarray}
Fluctuations of $u$ and $x$ (i.e., of the total population of the clouds and the population difference) are correlated in the broken-symmetry phase, with $\langle u(x-x_0)\rangle \approx (2x_0/N_c)[4  \theta + (W_{\rm out}/\tilde W)]^{-1}$, where $x_0=\pm(3  \theta)^{1/2}$.

In contrast, in the symmetric phase for $1\gg -  \theta\gg N_c^{-1/2}$ fluctuations of $u$ and $x$ are uncorrelated, with $N_c\sigma^2\approx |  \theta|^{-1} \left[1+(W_{\rm out}/2\tilde W)\right]\approx |\theta|^{-1}$.

It is seen from Eq.~(\ref{eq:variance_with_u_fluctuations}) that, for $  \theta \gg W_{\rm out}/4\tilde W$, the variance of the order parameter fluctuations in the broken-symmetry state scales with the control parameter in the same way as without fluctuations of the total population, $N_c\sigma^2\approx 5/4  \theta$. However, the factor in front of $1/\theta$ is now larger than in the symmetric phase. This corresponds to what was seen in the experiment. As $  \theta$ decreases below $W_{\rm out}/4\tilde W$, there occurs a crossover to $\sigma^2 \approx 1/2 N_c \theta$, i.e., $\sigma^2$  scales with $  \theta$ with the same scaling exponent, but with smaller amplitude than in the symmetric phase.

We note that there are other fluctuations that can contribute to the variance $\sigma^2$. For example, slow fluctuations of the trap parameters may lead to small and slow fluctuations of the total population, which can be thought of as fluctuations of the control parameter $\theta$. They give an extra contribution $3\langle(\delta\theta)^2\rangle/4\langle\theta\rangle$ to $\sigma^2$ in Eq.~(\ref{eq:population_fluct_symmetric_no_field}) for the symmetry-broken phase, where $\langle\delta\theta^2\rangle$ is the variance of $\theta$. Interestingly, this contribution also scales as $\theta^{-1}$ away from the critical point. It further increases the difference between the variance of the order parameter in the broken-symmetry and symmetric phase.

\section{One-dimensional model}
\noindent\subsection{The rotating wave approximation}

Calculation of the switching rates and response coefficients is simplified if the atomic motion in MOT is assumed one-dimensional, along the MOT axis $z$. This corresponds to averaging over the motion transverse to the axis, which is largely small-amplitude fluctuations about the vibrating centers of the clouds. In the experiment the radiation that confined the atoms transverse to the axis had large intensity, so that the correlation time of these fluctuations was short. In the single-atom approximation the atomic motion can be modeled by the Duffing oscillator,
\begin{eqnarray}
\label{eq:full_eom}
\ddot z + 2\Gamma \dot z +\omega_0^2(1+\epsilon_F\cos\omega_Ft)z +\gamma z^3= f(t)/m_a,
\end{eqnarray}
Here, $\omega_0$ and $\Gamma$ are the MOT eigenfrequency and viscous friction coefficient, respectively, $\gamma$ is the nonlinearity parameter, and $\epsilon_F$ is determined by the amplitude of the laser beam modulation \cite{Kim2005a}. These parameters are given in the main paper. The terms nonlinear in the velocity, which are disregarded in Eq.~(\ref{eq:full_eom}), are comparatively small for the experimental conditions. The function $f(t)$ is white thermal Gaussian noise, cf. Eq.~(\ref{eq:eom_oneatom_general}).

For the reduced modulation strength $\epsilon_F\omega_0/4\Gamma > 1$, resonant modulation ($|\omega_F-2\omega_0|\ll \omega_0$)  can excite period-two atomic vibrations \cite{LL_Mechanics2004}. The vibrations are nearly sinusoidal in the absence of noise, if the modulation is not too strong.  The atomic dynamics can be conveniently described by switching to the rotating frame using a standard transformation
\begin{eqnarray}
\label{eq: variable_transformation}
&z = C_{\rm RWA}\left[q_{2}\cos(\omega_F t/2)-q_{1}\sin(\omega_F t/2)\right],%\nonumber
\\
&\dot z = -C_{\rm RWA}(\omega_F/2)\left[q_{2}\sin(\omega_F t/2)+q_{1}\cos(\omega_F t/2)\right]
\nonumber
\end{eqnarray}
with $C_{\rm RWA}=(2\omega_0^2\epsilon_F/3\gamma)^{1/2}$ (the modulation sign is chosen in such a way that $\epsilon_F/\gamma > 0$). In the rotating wave approximation (RWA), the equations for slow variables $\qb\equiv (q_{1},q_{2})$ in slow time $\tau$ are
\begin{eqnarray}
\label{eq:eom_single_atom}
\frac{d\qb}{d\tau}={\bf K}^{}(\qb) + \fb(\tau), \qquad\tau=\omega_0\epsilon_Ft/4.
\end{eqnarray}
Here, $\fb(\tau)$ is white Gaussian noise with two asymptotically independent components, $\langle f_{\kappa}(\tau)f_{\kappa'}(\tau')\rangle = 2D_{\tau}k_BT\delta_{\kappa\kappa'}\delta(\tau-\tau')$, with $\kappa,\kappa'=1,2$ and $D_{\tau}=6\gamma\Gamma/m_a\omega_0^5\epsilon_F^2$. We use vector notations here {\it formally}, $\qb$ is not a vector in real space, its components are combinations of atomic displacement and momentum along the MOT axis.

Functions ${\bf K}^{}(\qb)$ are cubic polynomials and are given explicitly in Ref.~\onlinecite{Ryvkine2006a} [they were denoted by ${\bf K}^{(0)}(\qb)$]. The zeros of ${\bf K}$ occur at the stationary states in the rotating frame and, in turn, at the period-two vibrational states in the laboratory frame. We will be interested in the parameter range where Eq.~(\ref{eq:eom_single_atom}) has two attractors $\qb_1^{A}=-\qb_2^{A}$. Functions ${\bf K}^{}$ are also equal to zero for $\qb = {\bf 0}$, which corresponds to the unstable state of zero-amplitude period-two vibrations in the laboratory frame.

For low temperatures, the motion described by Eq.~(\ref{eq:eom_single_atom}) is primarily small-amplitude fluctuations about states  $\qb_{1,2}^{A}$. Switching between the states requires a large rare fluctuation. The single-atom switching rate has activation form $W^{(0)}=C\exp(-R^{(0)}/k_BT)$, Eq.~(\ref{eq:Wnm_general}), where as in Eq.~(\ref{eq:var_probl_R}),  $R^{(0)}=\min{\cal R}^{(0)}$ with
\begin{eqnarray}
\label{eq:switching_general}
{\cal R}^{(0)}&=&(4D_{\tau})^{-1}\int d\tau \fb^2(\tau) \nonumber\\
&&+\int d\tau{\bm\lambda}(\tau)\left[\frac{d\qb}{d\tau}-{\bf K}^{}-\fb(\tau)\right].
\end{eqnarray}
The major distinction from the general formulation (\ref{eq:var_probl_R}) is that function ${\bf K}^{}$ does not explicitly depend on time. This made it possible to solve the variational problem (\ref{eq:switching_general}) and to find both the optimal trajectories followed in switching from $n$th stable state $\qb_n^{\rm (opt)}(\tau)$ \cite{Dykman1998} and functions $\lab_n(\tau)$, which are equal to the logarithmic susceptibilities of Ref.~\onlinecite{Ryvkine2006a} multiplied by $-1/D_{\tau}$.

\noindent\subsection{Interatomic interaction}

The major inter-atomic interaction that affects the switching rate is the long-range interaction, since in switching atoms move far away from the clouds. It comes from the shadow effect, which is due to ``shielding"  of atoms from the laser light by other atoms \cite{Sesko1991,Steane1992,Dalibard1988,Walker1990,Mendoncca2008}. In the one-dimensional picture, the force on $i$th atom with coordinate $z^i$ from $j$th atom with coordinate $z^j$ can be approximately modeled as $F^{i}=-f_{\rm sh}\sum_j{\rm sgn}(z^{i}-z^j)$. This force is weak, much smaller than the Doppler force that confines atoms to the trap. Multiple scattering of light is disregarded in the above expression.

To estimate $f_{\rm sh}$ for a switching atom we have to take into account that the atomic clouds are 3-dimensional. We consider a simple model in which a laser beam with intensity $I$ propagating along $z$-axis passes on its way through an atomic cloud with density distribution $\rho(\rb)$. The resulting intensity change as a function of transverse coordinates $x,y$ is $\Delta I(x,y)=I\sigma_L\int dz \rho(\rb)$, where integration is done over the length of the cloud and $\sigma_L$ is the absorption cross-section. This cross-section depends in the standard way on the light intensity and the frequency detuning. Generally, because of the magnetic-field induced frequency shift and the Doppler shift $\sigma_L$ oscillates in time. The light intensity $I$ also oscillates in time. However, for comparatively small modulation amplitudes, we can disregard these oscillations. Then we get for typical experimental conditions $\sigma_L\approx 5.6\times 10^{-15}$~m$^2$.

The atomic density distribution $\rho(\rb)$ can be assumed Gaussian with the same width $w_t\approx 1$~mm in all directions, which was achieved in the experiment by tuning the transverse beam intensities. According to the optimal path picture, during switching atoms most likely move along the MOT axis, and for such atoms the light intensity change is determined by $\rho(\rb)$  on the axis. The extra force on the switching atom as it moves between the clouds is then directed toward the more populated cloud and is equal to $f_{\rm sh}|N_1-N_2|$, with
\begin{eqnarray}
\label{eq:shadow_force}
f_{\rm sh}=\hbar k \Gamma_p \sigma_L s/4\pi (s+1) w_t^2.
\end{eqnarray}
Here, $k$ is the photon wave number, $\Gamma_p$ is the reciprocal lifetime of the excited state, and $s=(I/I_s)[1+(2\delta/\Gamma_p)^2]^{-1}$ is the resonant absorption strength ($I_s$ is the saturation intensity and $\delta$ is the detuning of the radiation frequency from the atomic transition frequency). Equation (\ref{eq:shadow_force}) gives $f_{\rm sh}\approx 2.5\times 10^{-32}$~N. Note that, since the clouds are oscillating, the force $\propto f_{sh}$ is oscillating as well, its time dependence is determined by the $\sgn$-function.

\subsection{The shadow effect in the rotating frame}

The effect of the shadow force can be conveniently analyzed in the rotating frame by changing from $(z^{i},\dot z^{i})$ to slowly varying two-component vectors $\qb^{i}$, Eq.~(\ref{eq: variable_transformation}). In the RWA the equation of motion for $\qb^{i}$ becomes
\begin{eqnarray}
\label{eq:eom_interaction_no_noise}
\frac{d\qb^{i}}{d\tau}= {\bf K}^i(\qb^i)+  \hat\epsilon\partial_{\qb^{i}}H_{\rm sh} +\fb^i(\tau)
\end{eqnarray}
($i=1,\ldots,N_{\rm tot}$). Here, $\fb^i(\tau)$ is the random force on $i$th atom (random forces on different atoms are uncorrelated), $\hat\epsilon$ is the permutation tensor, and $H_{\rm sh}$ is the interaction Hamiltonian in the slow variables,
\begin{eqnarray}
\label{eq:H_sh}
&&H_{\rm sh}= \frac{1}{2}\sum^{\prime}\nolimits_{ij}V_{\rm sh}(\qb^{i}-\qb^{j}), \nonumber\\
&&V_{\rm sh}(\qb)=\frac{8f_{\rm sh}}{\pi m_a\omega_0^2\epsilon_F C_{\rm RWA}}|\qb|,
\end{eqnarray}
where the prime indicates that $i\neq j$.

%\noindent\subsubsection{1. Many-atomic switching rate}

The general expression for the interaction-induced correction to the activation energy $R_n$, Eq.~(\ref{eq:many_atom_correct_to_Rn}), in the RWA has the form
\begin{eqnarray}
\label{eq:alpha_RWA}
\alpha_{nm}=\int_{-\infty}^{\infty}d\tau\lab_n(\tau)
\hat\epsilon\partial_{\qb_m^{A}}V_{\rm sh}(\qb_n^{\rm (opt)}(\tau)-\qb_m^{A}),
%\nonumber\\
%\n_m\equiv \partial_{\qb_m^{A}},
\end{eqnarray}
where $\lab_n(\tau)$ and $\qb_n^{\rm (opt)}(\tau)$ are the solutions of the single-atom problem of minimizing the functional (\ref{eq:switching_general}) for switching from the state $n$.

Equations (\ref{eq:switching_general}) and (\ref{eq:alpha_RWA}) were used to find the coefficients $\alpha_{nm}$ and the critical value of the total number of atoms $N_c=2k_BT/(\alpha_{11}-\alpha_{12})$ where the symmetry breaking transition occurs. The obtained $N_c$ was within a factor of $2$ from the value of $N_c$ observed in the experiment. This is reasonable given the uncertainty in the temperature, the renormalization of $N_c$ by the finite lifetime of atoms in the MOT, cf. Eq.~(\ref{eq:theta_renormalization}), and the fact that the light intensity modulation was not weak and, respectively, the atomic vibrations noticeably deviated from sinusoidal.

An important effect that allows one to independently compare the many-atomic theory with experiment is the change of the vibration amplitudes due to the interaction. This change is determined by the change of the equilibrium positions $\qb_n^{A}$ in the rotating frame. For weak inter-atomic interaction it can be found by linearizing equations of motion (\ref{eq:eom_interaction_no_noise}). The average force on an atom from other atoms in the same cloud is equal to zero. The shift of the equilibrium position is due only to the force from the atoms in the other cloud. Disregarding small fluctuations about the equilibrium positions, for the shift of $n$th attractor in the rotating frame $\delta \qb_n^{A}$ we obtain
\begin{eqnarray}
\label{eq:attractor_shift_more_epxlicit}
(\delta\qb_n^{A}\partial_{\qb_n^{A}}){\bf K}^{}\left(\qb_n^{A}\right) +N_{3-n}\hat{\epsilon}\partial_{\qb_n^{A}}\bar V_{\rm sh}(\qb_n^{A}-\qb_{3-n}^{A})=0.
\end{eqnarray}
Here, $\bar V_{\rm sh}$ is given by Eq.~(\ref{eq:H_sh}) for $V_{\rm sh}$ in which $f_{\rm sh}$ is replaced with $\bar f_{\rm sh}$. The quantity $\bar f_{\rm sh}$ characterizes the force from the shadow effect which is averaged over the cross-section of the cloud transverse to the MOT axis. Such averaging occurs as atoms are moving within the cloud. For Gaussian clouds of the same cross-section, $\bar f_{\rm sh}=f_{\rm sh}/2$.

The shift of the $n$th attractor ($n=1,2$) given by Eq.~(\ref{eq:attractor_shift_more_epxlicit}) and the corresponding change of the vibration amplitude of the $n$th cloud is proportional to the number of atoms in the other cloud, $N_{3-n}$. Therefore the vibration amplitudes of the clouds become different in the broken-symmetry state. This prediction and the linear dependence of the amplitude change on the number of atoms in the other cloud are in full agreement with the experiment; the numerical estimate for the simplified model of sinusoidal 1D vibrations is within a factor of 2 from the measured value.

%\addtocounter{bibliography}{20}

%\bibliographystyle{apsrev}
%\bibliography{C:/Aaa/BibTex/md10}